\documentclass[showpacs,amsmath,amssymb,prd]{revtex4}
\usepackage{graphicx}
\usepackage{dcolumn}
\usepackage{bm}
\usepackage{epsfig}
\usepackage{multirow}
\usepackage[a4paper]{geometry}
\usepackage[in]{fullpage}
\usepackage{color}
\newcommand{\beq}{\begin{equation}}
\newcommand{\eeq}{\end{equation}}
\newcommand{\beqarray}{\begin{eqnarray}}
\newcommand{\eeqarray}{\end{eqnarray}}

\begin{document}

\title{Coalescence Time of Compact Binary Systems}

\author{Lutendo Nyadzani} 
\email{lnyadzani@uj.ac.za} 
\author{Soebur Razzaque}
\email{srazzaque@uj.ac.za}

\affiliation{Centre for Astro-Particle Physics (CAPP) and Department of Physics, University of Johannesburg, Auckland Park 2006, South Africa}

\begin{abstract}
Compact binary systems coalesce over time due to the radiation of gravitational waves, following the field equations of general relativity. Conservation of energy and angular momentum gives a mathematical description for the evolution of separation between the orbiting objects and eccentricity ($e$) of the orbit. We develop an improved analytical solution to the coalescence time for a circular binary system with an arbitrary semi-major axis, to the first post-Newtonian (1PN) accuracy. The results from the quadruple approximation and 1PN approximation are compared for a circular orbit ($e=0$).
\end{abstract}

\date{\today}
\maketitle 	

%
%
\section{Introduction}
Compact objects are stellar remnants, such as black holes (BHs), neutron stars (NSs) and white dwarfs (WDs). A compact binary system consists of two compact objects (BH-BH, BH-NS, BH-WD, NS-NS, NS-WD) orbiting about a common center of gravity. According to the general theory of relativity (GR) these systems radiate energy in the form of gravitational waves (GWs)~\cite{einstein1918gravitationswellen, einstein1916naherungsweise}. The radiation of GWs results in the orbital properties of the system changing with time and eventually leading to coalescence of the two objects. 

The first indirect evidence of GW radiation was found in a binary pulsar system discovered by Hulse and Taylor, well known as the Hulse-Taylor binary~\cite{Hulse:1974eb, Taylor:1982zz}. On 14 September 2015, the first direct detection of GWs was achieved by the advanced Laser Interferometer Gravitational Wave Observatory (LIGO)~\cite{abbott2016observation}. This has been followed by detections of GWs by LIGO and Virgo from GW151012, GW151226, GW170104, GW170729, GW170814, GW170817, GW170818, GW170823, etc.~\cite{ligo2018gwtc}. These detections started a new era in astronomy and astrophysics. Among the 10 GW signals detected by LIGO and Virgo during the first and second observation runs, 9 were binary BH-BH mergers~\cite{ligo2018gwtc} and GW170817 was a binary NS-NS merger~\cite{abbott2017gw170817, ligo2018gwtc}. The NS-NS merger event was also followed by electromagnetic detection~\cite{Goldstein:2017mmi, GBM:2017lvd, Troja:2017nqp, DAvanzo:2018zyz, Resmi:2018wuc} of a short gamma-ray burst (GRB), which confirmed the theory~\cite{eichler1989nucleosynthesis, Narayan1992iy, metzger2010electromagnetic} that at least some short GRBs are produced by the coalescence of two NSs~\cite{abbott2017gw170817}.

With these observations, it is now more important to study the evolution and dynamics of compact binary systems. The orbital evolution of compact binaries and the radiation of GWs have been studied in detail since their first prediction by Einstein. However, to study binary systems in GR one must solve the complicated Einstein field equations. These equations have no known general solution, only approximations and special case solutions are known. One of the commonly used approximation is the weak field approximation, where one linearizes the Einstein field equations. This approach was used by Einstein himself to predict the existence of GW ~\cite{einstein1918gravitationswellen, einstein1916naherungsweise}. Following the same approach as Einstein, Peters studied the orbital dynamics of binary systems in quadruple approximation in detail in Ref.~\cite{peters1964gravitational, peters1963gravitational}. In Ref.~\cite{peters1964gravitational} it is shown that the orbit will shrink and circularize due to the emission of GW and the coalescence time scale was calculated in the quadruple approximation.

In this paper, we present the coalescence time scale of a compact binary system with circular orbit ($e=0$) and arbitrary semi-major axis ($a$) in the first post-Newtonian approximation (1PN). We begin by reviewing the analytical and numerical solutions done by Peters~\cite{peters1964gravitational} in the quadruple approximation, and then present our analytical solution of the coalescence time in the 1PN approximation for a circular orbit. Finally, we compare the coalescence time scales in the quadruple and 1PN approximations.  
%
%
\section{Evolution of compact binary system emitting GWs}
The energy and angular momentum losses due to GW radiation by a compact binary system of masses $m_1$ and $m_2$ in a close orbit around a common centre of gravity with eccentricity $e$ and semi-major axis $a$ are given in the quadruple approximation  by~\cite{peters1964gravitational, peters1963gravitational} 
\begin{align}
\Big\langle\frac{dE}{dt}\Big\rangle = -\frac{32}{5}\frac{G^4m_1m_2(m_1+m_2)}{c^5a^5(1-e^2)^{\frac{7}{2}}} \Big(1+\frac{73}{24}e^2 +\frac{37}{96}e^4\Big)
\label{eq:E_loss}
\end{align}
\begin{align}
\Big\langle\frac{dL}{dt}\Big\rangle = -\frac{32}{5}\frac{G^{\frac{7}{2}}m_1^2m_2^2(m_1+m_2)^{\frac{1}{2}}}{c^5a^{\frac{7}{2}}(1-e^2)^{2}} \Big(1+\frac{7}{8}e^2\Big)
\label{eq:L_loss}
\end{align}
where $G$ is the universal gravitational constant and $c$ is the speed of light. Equations~(\ref{eq:E_loss}) and (\ref{eq:L_loss}) are averaged over the orbital period. The change in $e$ and $a$ are found by equating the gravitational energy of the two body problem, $E = -G m_1 m_2/2a$ into equation~(\ref{eq:E_loss}) and the corresponding angular momentum, $L^2 = G m_1^2 m_2^2a(1-e^2)/(m_1+m_2)$ into equation~(\ref{eq:L_loss}) as 
\begin{align}
\frac{da}{dt}=\frac{-\beta}{a^3(1-e^2)^{\frac{7}{2}}}
\Big(1+\frac{73}{24}e^2+\frac{37}{96}e^4\Big)
\label{da}
\end{align}
\begin{align}        
\frac{de}{dt}=-\frac{19}{12}\frac{\beta}{a^4(1-e^2)^{\frac{5}{2}}}
\Big(e+\frac{121}{304}e^3\Big)
\label{de}
\end{align}
where 
\begin{align}
\beta = \frac{64G^3m_1m_2(m_1+m_2)}{5c^5}
\label{beta}
\end{align}
%
%
\section{Coalescence time}
An analytical solution to the coalescence time in the quadruple approximation and in the case of a circular orbit $(e=0)$, is given by~\cite{peters1964gravitational}, 
\begin{align}
T_c(a_0)=\frac{a^4_0}{4\beta}\label{Tc}
\end{align}
where $a_0$ is the semi-major axis at the time of the observation. In the general case, with arbitrary values of $a$ and $e$, the coalescence time can be calculated by solving the coupled  equations (\ref{da}) and (\ref{de}) together. In particular, dividing equation (\ref{da}) by (\ref{de}) and subsequently integrating over $e$ gives $a$ as a function of $e$, which is given by
\begin{align}
a(e)=\frac{k_0e^{\frac{12}{19}}}{(1-e^2)}\Big[1+\frac{121}{304}e^2\Big]^{\frac{870}{2299}}\label{a}
\end{align}
Here $k_0$ is a function of the initial conditions $a=a_0$ and $e=e_0$. By substituting equation (\ref{a}) into equation (\ref{de}) and integrating with respect to $e$ and $t$ gives the following coalescence time~\cite{peters1964gravitational}
\begin{align}
T(a_0,e_0) = \frac{12k_0^4}{19\beta} \int ^{e_0}_0{ \frac{e^{\frac{29}{19}} \left( 1+ \frac{121}{304} e^2 \right)^{\frac{1181}{2299}}}{(1-e^2)^{\frac{3}{2}}}de}
\label{T}
\end{align}
in the quadruple approximation. This integral has no simple analytical solution\footnote{See, however, Ref.~\cite{Berry:2013ara} for a solution using hypergeometric functions.}, but it can  be solved numerically. 
%
%
\section{Coalescence time to first post-Newtonian accuracy}
Following the same argument as in Ref.~\cite{peters1964gravitational}, that a massive binary system will coalesce due to the emission  of GW, we compute the coalescence time in the first post-Newtonian approximation.  
In the 1PN order, the orbit shape of the semi-major axis and the eccentricity, $a_1$ and $e_1$, are given by \cite{AIHPA19854311070},
\begin{align}
a_1=-\frac{GM}{2E_1}-\frac{GM(7-\nu)}{4c^2}
\label{a1}
\end{align}
\begin{align}
e_1=1+2E_1 h_1^2+\frac{E_1[E_1h_1^2(3-\nu)+2(6-\nu)]}{c^2}
\label{e1}
\end{align}
Here $\nu$ is given by $\nu=\mu/M$ with $M=m_1+m_2$ and $\mu$ is the reduced mass given by $\mu=m_1m_2/(m_1+m_2)$. The variables $E_1$ and $h_1$ are the 1PN reduced energy and reduced angular momentum, respectively.  In this order the average GW luminosity is given by \cite{Blanchet:1989cu},
\begin{align}
\notag
\big<\cal{L}\big>=&\frac{1024 \nu^2(-E_1)^5}{4Gc^5(1-e_1^2)^{7/2}}\Big[1+\frac{73}{24}e_1^2+\frac{37}{96}e_1^4-\frac{E_1}{168c^2(1-e_1^2)}\\&\times[13-6414e_1^2-\frac{27405}{4}e_1^4-\frac{5377}{4}e_1^6+\nu(-840-\frac{6419}{2}e_1^2-\frac{5103}{8}e_1^4+\frac{259}{8}e_1^6)]\Big]\label{LM}
\end{align}
Equation~(\ref{LM}) is analogous to equation~(\ref{eq:E_loss}) with the fourth term in equation~(\ref{LM}) being the 1PN correction, and $E_1$ is the 1PN reduced energy given by \cite{AIHPA19854311070}, 
\begin{align}
E_1=-\frac{GM}{2a_1}+\frac{GM}{8c^2a_1^2}(7-\nu)\label{Ea}
\end{align}
Equation~(\ref{eq:E_loss}) can be obtained from equation~(\ref{LM}) by letting $E_1$ equal to the Newtonian energy $E_1 = E_N = GM/2a_N$, here $a_N$ is the Newtonian semi-major axis, and by letting $e_1$ equal to the Keplerian eccentricity $e$.

The rate of energy loss due  to the emission of GW is 
\begin{align}
\frac{dE}{dt}=- \frac{\big<\cal{L}\big>}{\mu}
\label{dEdt}
\end{align}
The time scale of a system emitting GW is given by 
\begin{align}
T=-\mu \int^{E^\prime}_{E_0}{\frac{1}{\big<\cal{L}\big>}dE}
\label{eqL}
\end{align}
This is the time it will take for the energy of the system to change from its initial value $E_0(a_0,e_0)$ to $E^\prime(a^\prime, e^\prime)$. This integral has no general analytical  solution for any arbitrary value of $e_1$. However, an analytical solution can be found for the circular orbit by setting $e_1=0$ and letting $E_1=E_N$ in equation~(\ref{LM}). Subsequently we obtain the equation for $da_1/dt$ as
\begin{align}
\Big<\frac{da_1}{dt}\Big>=\frac{\beta_0(a_1-\beta_1)^5(a_1^2+\beta_2a_1-\beta_1\beta_2)}{a_1^9(a_1-2\beta_1)} \label{da1}
\end{align}
Here $\beta_0=\beta$ as in equation~(\ref{beta}), $\beta_1=GM(7-\nu)/(4c^2)$ and $\beta_2 = GM(13-840\nu)/(336c^2)$. This equation describes how the semi-major axis changes with time due to the emission of GW. Equation~(\ref{da}) for the quadruple approximation can be recovered from equation~(\ref{da1}) by neglecting the terms with $\beta_1$ and $\beta_2$. Solving the integral in equation~(\ref{eqL}) with $e_1=0$ and $E_1$ given by equation~(\ref{Ea}) we obtain the following coalescence time 
\begin{align}
\notag
T_{c,1PN}(a_1)=& \frac{a_1^4}{4\beta}+\frac{a_1^3C_1}{3\beta} +\frac{a_1^2C_2}{2\beta} +\frac{a_1C_3}{\beta}+\frac{\beta_2^4}{\beta} \log\Big(\frac{a_1^2+a_1 \beta_2-\beta_1\beta_2}{a_1-\beta_1}\Big)+\frac{\beta_1^8}{4 \beta(a_1-\beta_1)^4}\\
&+\frac{C_4}{3\beta (a_1-\beta_1)^3}+\frac{C_5}{2 \beta(a_1-\beta_1)^2}+\frac{C_6}{\beta(a_1-\beta_1)}\label{T1PN}
\end{align}
where 
\begin{align}
\notag
C_1 = &3\beta_1-\beta_2\\
\notag
C_2 = & 5\beta_1^2-2\beta_1\beta_2+\beta_2^2\\
\notag
C_3 = &5\beta_1^3-2\beta_1^2\beta_2+\beta_1\beta_2^2-\beta_2^3\\
\notag
C_4 = &\beta_1^6(6\beta_1-\beta_2)\\
\notag
C_5= & \beta_1^4(14\beta_1^2-4\beta_1\beta_2+\beta_2^2)\\
\notag
C_6 = & \beta_1^2(14\beta_1^3-5\beta_1^2\beta_2+2\beta_1\beta_2^2-\beta_2^3)
\end{align}
and $\beta$ has the same meaning as in equation (\ref{beta}). The semi-major axis $a_1$ in  equation (\ref{T1PN}) is the post-Newtonian semi-major axis given by equation (\ref{a1}).  Letting the energy $E_1$ in equation (\ref{a1}) be equal to the Newtonian energy $$E_1=\frac{GM}{2a_N}$$ we find that the 1PN orbit is smaller than the Newtonian orbit.  However letting the semi-major axis in equation~(\ref{Ea}) be equal to the Newtonian semi-major axis it can be shown that the post-Newtonian orbit has more energy compared to the Newtonian orbit.
 
\begin{figure}[h]
\centering
\includegraphics[width=10cm]{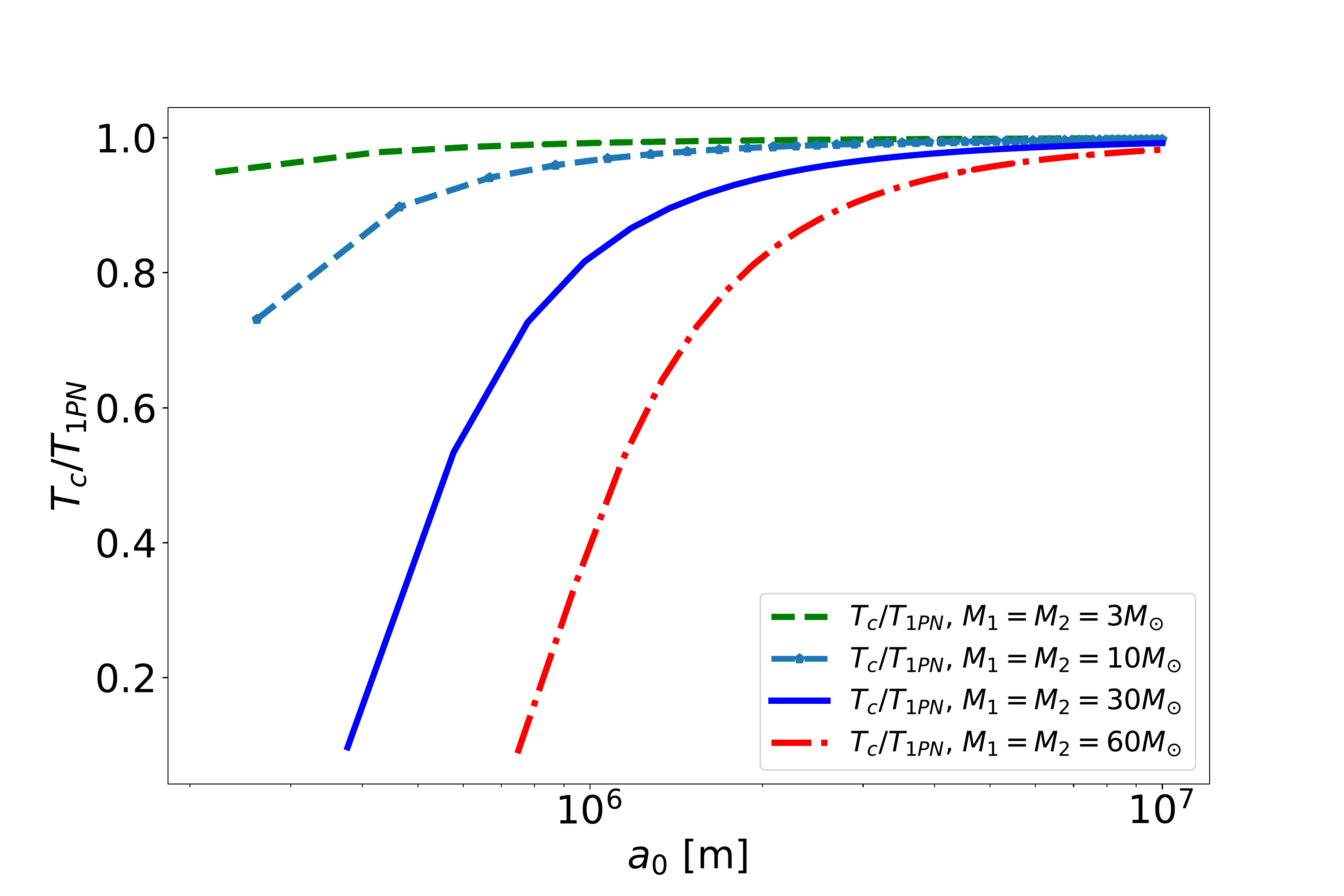}
\caption{Ratio of the coalescence time scales in the quadruple approximation ($T_c$) to the 1PN approximation ($T_{1PN}$) versus the initial semi-major axis ($a_0$) for a circular binary systems emitting GWs.  Different lines are for different (equal) masses ($3M_{\odot}$, $10M_\odot$, $30M_{\odot}$ and $60 M_{\odot}$) in the binary. It is clear that the quadruple approximation in equation~(\ref{Tc}) increasingly underestimates the coalescence time compared to the 1PN approximation in equation~(\ref{T1PN}) with increasing binary mass and decreasing semi-major axis.}
\label{fig3}
\end{figure}

Figure \ref{fig3} shows the ratio between the time scale from the quadruple approximation $(T_c)$ in Ref. ~\cite{peters1964gravitational} and our first post-Newtonian time scale $(T_{c1PN})$. In order to make this comparison we let the energy $E_1$ in equation (\ref{a1}) be equal to the Newtonian energy,
$$E_1=E_N$$ such that $$a_1=a_N-\beta_1$$  
the semi-major axis is ranging from the initial semi-major axis to the Schwarzschild radius, $R_s=2GM/c^3$ of this orbit. These results show that the time-scale in quadruple approximation~\cite{peters1964gravitational} underestimates the actual time-scale in the first post-Newtonian accuracy. Hence the 1PN orbits decay slower than the quadruple approximation predicted in Ref.~\cite{peters1964gravitational}. The difference between $T_c$ and $T_{c1PN}$ increases with decreasing semi-major axis and increasing mass of the compact object. This is expected since the 1PN correction becomes more impotent as the strength of the gravitational field increases. Figure~\ref{fig3} also shows that the 1PN correction affect the GW time-scale decay effectively as the semi-major axis comes close to the Schwarzschild radius.

These calculations are done for a circular orbit. For a general orbit with arbitrary eccentricity and semi-major axis, one needs to solve equation~(\ref{eqL}) for every $0<e_1<1$.  This can be done by writing $e_1$ as a function of $a_1$ in equation~(\ref{e1}), and the results will be presented elsewhere. 
%
%

\section{Summary}
In summary, we have provided an analytical solution to the coalescence time for compact binary systems emitting gravitational waves in the first post-Newtonian approximation.  Our results show that the quadruple approximation underestimates the coalescence time scale of a compact binary system.  The difference with our 1PN time scale is significant during the final orbits of the binary system and for more massive objects. Our results have implications for compact binary merger rate calculations for different astrophysical source populations. 

\acknowledgments
This work was supported in part by the National Research Foundation (NRF, South Africa) with Grant No: 111749 (CPRR) to SR. LN acknowledges support from an NRF MSc student bursary.

\bibliography{Reference}  

\end{document}